\input psfig.tex
\documentstyle{article}
\begin{document}
{\pagestyle{empty}
\font\fone=cmr10 scaled\magstep3
\font\ftwo=cmr7 scaled\magstep3
{\topskip 0.5in
\vskip 1.5in
\centerline{\fone Spontaneous Symmetry Breaking in the SO(3)}
\centerline{\fone Gauge Theory to Discrete Subgroups}
\vskip0.3in
\centerline{\ftwo G\'abor Etesi} 
\vskip0.1in
\centerline{Institute for Theoretical Physics}
\centerline{E\"otv\"os Lor\'and University}
\centerline{Puskin u. 5-7, Budapest}
\centerline{H-1088 Hungary} 
\centerline{{\tt etesi@hal9000.elte.hu}} 
\vskip0.3 in
\begin{abstract}
In this paper we give a systematical description of the possible symmetry
breakings in the $SO(3)$-gauge theory and show an algorithmical method
to construct $SU(2)$- or $SO(3)$-invariant Higgs potentials in an arbitrary irreducible representation using regular graphs. We close our paper with the 
explicit construction of the Lagrangian of the simplest $SO(3)\rightarrow A_4$
theory.
\end{abstract}
\centerline{PACS 11.15.-q}
\centerline{PACS 11.15.Ex}
{\pagestyle{myheadings}
\markright{G. Etesi: Spontaneous symmetry breaking} 
\section{Introduction}
A very interesting area of the today's theoretical physics is the study of the
so-called discrete gauge theories (discrete Yang-Mills theories)[1],[2],[3]. A familiar
way to construct such theories is to break down the continuous symmetry of a
usual gauge theory using Higgs mechanism. But if we want to give an explicit
example of such a symmetry breaking we need to solve two non trivial problems.

Firstly, how could we produce invariant polynomials of the initial gauge group
in an arbitrary representation? This is a very hard problem of theory of group
invariants and we cannot answer the question generally even in the very
simple case of $SU(2)$.

Our second problem is to find a representation of the initial gauge group $G$
for a given subgroup $H\subset G$ which possesses the symmetry breaking $G\rightarrow
H$. Generally this is an algebro-geometrical question, because we can equivalently
say that we must find points on the zero variety of the $G$-invariant polynomial
having given stabilizer subgroup $H$ under the action of $G$.

In the case of the group $SU(2)$ we were able to develop a simple method
using regular graphs to make $SU(2)$-(and of course $SO(3)$-) invariant
polynomials.
\newpage}
{\pagestyle{myheadings}
\markright{G. Etesi: Spontaneous symmetry breaking}
Because the subgroups of the group $SO(3)$ have contacts with regular two and
three dimensional polyhedra, using simple methods from the theory of group
representations we can list all possible stabilizer subgroups in an arbitrary
irreducible representation of the group $SO(3)$.           
\section{Statement of the problem}
If we want to break down the symmetry in a given gauge theory with gauge group $G$
we need to give explicitly a so-called Higgs-potential $V$ which is a polynomial
in the Higgs scalar fields and satisfies the following conditions:
\begin{itemize}
\item $V$ is invariant under the action of the group $G$ in a given
representation;

\item $V$ is bounded; 

\item the self interactions of the scalar fields induced by the polynomial $V$
are renormalizable.
\end{itemize}

If we want to break the gauge symmetry to a given subgroup of the gauge group $G$ we need some more information about the polynomial $V$. Our starting point is
the familiar Lagrangian:
$${\cal L}=-{1\over 4}F^{\mu\nu}_aF^{a}_{\mu\nu}+(D^{\mu}\Phi)^{\star}(D_{\mu}
\Phi)-V(\Phi).\eqno (1)$$
Here $V: {\bf k}^n\rightarrow {\bf R}$ is a polynomial satisfies the above 
properties (${\bf k}$ denotes ${\bf C}$ or ${\bf R}$). We can fix the minimum
of the polynomial $V$ to be zero. Let $Z(V=0)$ denote the zero variety of $V$;
so a vacuum state of the theory is given by $A_{\mu}^a=0$ and $\Phi=\Phi _0\in
 Z(V=0)$. Using the potential $V$ we can ``break the symmetry spontaneously
down'' which means that the vacuum state (which is a point in $Z(V=0)$)
has no more the whole dynamical symmetry (the group $G$) but a subgroup $H$ of
$G$ only. This subgroup $H$ has the property that its elements stabilize the 
vacuum state (i. e. $H\Phi _0=\Phi_0$). So this subgroup $H$ is the {\it stabilizer
subgroup} of the point $\Phi_0\in Z(V=0)$. We are interested in such situations when this group is {\it discrete}. Now we are in position to give a precise formulation
of our problem.

 Let us consider the field theory (1) with symmetry group $G$ 
(We assume that this group is an algebraic subgroup of some $GL({\bf k}^m)$)
and let us take a discrete subgroup of it. Also take a representation $\rho : G\rightarrow GL({\bf k}^n)$ of the group $G$. We are searching for polynomials $V:
{\bf k}^n\rightarrow {\bf R}$ which satisfy the following conditions:
\begin{itemize}
\item $V$ is invariant under the action $\rho$ of $G$ on ${\bf k}^n$;

\item $V$ is bounded;

\item and $Z(V=0)$ has a subset with stabilizer subgroup $H\subset G$.
\end{itemize}
We have omitted the condition of renormalizability because this is a simple 
restriction of the degree of the polynomial $V$.

In the case $G=SO(3)$ we can solve the problem generally: we are able to list all possible symmetry violation and can show a simple algorithmical method to
construct Higgs potentials in arbitrary high dimensional Higgs representations.
Let us see how to do this!
\section{Construction of invariant polynomials}
Let $j$ be an integer or half-integer number and let us take the space of all homogeneous 
complex polynomials $p_{2j}(x,y)$ having two variables and homogeneous degree
$2j$. This space is naturally identified with ${\bf C}^{2j+1}$. By the aid of
the canonical two dimensional representation of $SU(2)$ we can describe  a
$2j+1$ dimensional representations as follows. If $\left(\matrix{\alpha & \beta
\cr -\overline{\beta} & \overline{\alpha}\cr}\right)\in SU(2)$ the transformation of a vector
$(x,y)\in {\bf C}^2$ given by
$$x\rightarrow \alpha x+\beta y,$$
$$y\rightarrow -\overline{\beta}x+\overline{\alpha}y.$$
Using this equations we get the transformation rule of a homogeneous polynomial: $$\sum\limits_{n=0}^{2j}a_nx^ny^{2j-n}\rightarrow \sum\limits_{n=0}^{2j}a_n
(\alpha x+\beta y)^n(-\overline{\beta}x+\overline{\alpha}y)^{2j-n}.$$
Clearly this is a $2j+1$ dimensional representation of the group $SU(2)$.
The irreducibility of this representation is due to the fact that in the two
dimensional representation of the $SU(2)$ there are not $SU(2)$-invariant
polynomials. Let $\lambda_n$ denote the roots of the polynomial $p_{2j}(x,1)$.
Now we can write
$$\sum\limits_{n=0}^{2j}a_nx^ny^{2j-n}=a_{2j}\prod\limits_{n=1}^{2j}(x-\lambda_ny).$$
{\bf Definition}. {\it Let $k$,$l$ positive integer numbers The graph ${\cal G}$
is called {\bf $k$-regular oriented graph of order $l$} if it has $l$ vertices, in every vertex converge $k$ edges and every edge has an orientation}.
\vskip 0.2 in
For example {\bf Figure 1} shows a 4-regular oriented graph of order 5.
\vskip 0.1in
\centerline{\psfig{figure=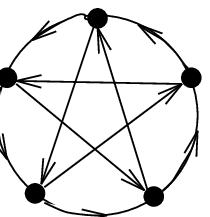,height=0.8in}}
\vskip 0.081in
\centerline{{\bf Figure 1.}}
\centerline{4-regular graph of order 5.}
\vskip 0.1in 
Now let ${\cal G}$ be a $k$-regular oriented graph of order $2j$. We order to
${\cal G}$ an expression:
$$a_{2j}^k\prod\limits_{(mn)}(\lambda_m-\lambda_n).\eqno (2)$$
Here the product $\prod\limits_{(mn)}$ is understood as the product (2) has to
involve a factor $\lambda_m-\lambda_n$ if in its graph there is an edge of
the form seen on {\bf Figure 2}.
\vskip 0.1in
\centerline{\psfig{figure=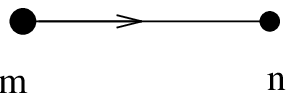,height=0.3in}}
\vskip 0.081in
\centerline{{\bf Figure 2.}}
\centerline{A typical part of a graph.}
\vskip 0.1in
(We can connect two vertices with more than one edges, in this situation
we count the edges with multiplicity.)

On a graph of order $2j$ the permutation group $S_{2j}$ acts naturally 
transposing the vertices of the graph. This gives the symmetrization of the
expression (2).
\vskip0.2 in
{\bf Proposition 1.} {\it The expression
$${1\over\vert S_{2j}\vert}a^k_{2j}\sum\limits_{\pi\in S_{2j}}\prod\limits_{(mn)}(\lambda_{\pi (m)}-\lambda_{\pi (n)})\eqno (3)$$
is invariant under the action of $SU(2)$}.$\diamondsuit$
\vskip 0.2in
Let $\sigma_n$ denote the $n^{th}$ elementary symmetric polynomial with 
variables $\lambda_1,...,\lambda_{2j}$. The expression (3) clearly symmetric in
$\lambda_1,...,\lambda_{2j}$ so it is uniqly expressible as a polynomial in
$\sigma_1,...,\sigma_{2j}$. But using the well-known relations between the
roots and coefficients of a polynomial, $\sigma_n=(-1)^n{a_{2j-n}\over a_{2j}}$,
we have the result that (3) is an $SU(2)$-(or, if $j$ is an integer an $SO(3)$-) invariant polynomial of the form $f(a_0,...,a_{2j})$ of homogeneous degree $k$.

The easy proof is left to the reader.

One can use this method very effectively if one has a computer (because of the
symmetrization of the expression (2)). We have computed some invariants of
$SU(2)$. Now we show only the 7 dimensional invariant of degree 2 illustrated
on {\bf Figure 3}.
\vskip 0.1in
\centerline{\psfig{figure=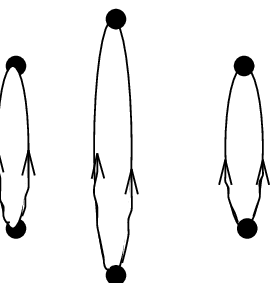,height=0.8in}}
\vskip 0.81in
\centerline{{\bf Figure 3.}}
\centerline{$6a_3^2-16a_2a_4+40a_1a_5-240a_0a_6$}
\centerline{2-regular graph of order 6 and its $SU(2)$-invariant polynomial}
\section{Classification of symmetry violations of the $SO(3)$-theory}
Now we turn to our second problem to classify all possible symmetry breaking in the 
$SO(3)$-gauge theory. Let $G\subset SO(3)$ a subgroup and its trivial representation
given by $g\rightarrow 1\in GL(1,{\bf R})={\bf R}^*$, $g\in G$. We say that 
$G\subset SO(3)$ is a maximal subgroup of the group $SO(3)$ if there is no
subgroup $H\subset SO(3)$ satisfying $G\subset H$. The clue of the description is the
following simple proposition.
\vskip 0.1in
{\bf Proposition 2}. {\it Let $\rho : SO(3)\rightarrow GL(V)$ be an irreducible
representation of the group $SO(3)$ and let $G$ its maximal subgroup. If the
direct decomposition of $\rho$ according to $G$ contains the trivial representation 
of $G$ then in $V$ there is a subspace $W_G$ whose points are stabilized by the
group $G$. Moreover the dimension of $W_G$ is equal to the multiplicity of the
trivial representation of $G$ in $\rho$}.$\diamondsuit$
\vskip 0.1in
The straightforward verification of proposition 2 is left to the reader.
So if the subgroup $G$ satisfies the condition of proposition 2 the only thing
what we need to do is to determine the characters of the group $G$ in the
representation $\rho$. But this is not difficult. Firstly if we take a review
about the (discrete) subgroups of $SO(3)$ we find that these groups are closely
related to well-known geometrical objects: these groups are the symmetry groups
of the two and three dimensional regular polyhedra. If we take into account
this fact we are able to construct these groups as a set of rotations under
which
the adequate regular polyhedron remains invariant (but not pointwise).
   
But if we know these rotations we can easily give the characters of the subgroup $G$ in the representation $\rho$ since
$$\chi_j(\phi)={\sin(j+{1\over 2})\phi\over\sin{\phi\over 2}}.$$
Here $j$ denotes the weight of the representation $\rho$. 

If the group is not maximal i.e. exists a subgroup $H$ such that $G\subset H
\subset SO(3)$, then we need to consider the multiplicity of the trivial representation of $G$ respectively of $H$.
If the multiplicity of the trivial $G$-representation is bigger than the 
multiplicity of the trivial $H$-representation then there are points in $V$
whose stabilizer subgroup is the not-maximal subgroup $G$. Leaving some
technical details we get in summary the table at the end of the paper.
\section{The $SO(3)\rightarrow A_4$ theory}
The time has come to examine explicitly a not usual symmetry violation.
Using the Table at the end of the paper we can see: it is possible to violate
the $SO(3)$ gauge symmetry to its non-Abelian subgroup $A_4$ using 7 dimensional
Higgs representation. We choose for this procedure the potential showed on
{\bf Figure 3}. The 
7 dimensional representation is constructed by the above polynomial method and
is the 7 dimensional complex irreducible representation of the group $SO(3)$, too.  
Firstly we need to find a 7 dimensional {\it real} irreducible subspace 
${\bf R}^7$ of ${\bf C}^7$ which gives the {\it real} representation of the 
$SO(3)$. Not difficult to see that a simple basis of this real subspace is
given by the polynomials which satisfy the functional equation
$$p(x,y)=-\overline{p}(-\overline{y},\overline{x}).$$
If we want to get an orthogonal real representation we need to multiply these
vectors by certain numerical factors and get:
$${i\over\sqrt{120}}(x^6+y^6);\quad {1\over\sqrt{120}}(x^6-y^6);$$
$${1\over\sqrt{20}}(x^5y+xy^5);\quad {i\over\sqrt{20}}(x^5y-xy^5);$$
$${i\over\sqrt{8}}(x^4y^2+x^2y^4);\quad{1\over\sqrt{8}}(x^4y^2-x^2y^4);\eqno (5)$$
$${1\over\sqrt{3}}x^3y^3.$$
In this basis the polynomial illustrated on {\bf Figure 3} has the simple form
$$a_0^2+a_1^2+a_2^2+a_3^2+a_4^2+a_5^2+a_6^2$$
which shows that this polynomial is bounded. Now we turn to our next question:
how to find the coordinates of the $A_4$ vacuum?
The group $A_4$, the $4^{th}$ alternating group, has two generators denoted
by $a$, $b$. Clearly, our points need to be in the linear space
${\bf R}^7\cap{\bf Ker}(\rho (a)-Id)\cap{\bf Ker}(\rho (b)-Id)$ and have to
be normed. After constructing the 7 dimensional representation of $a$ and $b$
we get the two possible vacuum states in the basis (5):
$$\pm {1\over\sqrt{270}}\left(\matrix{\sqrt{120}\cr 0\cr 0\cr 5\sqrt{6}\cr 0
\cr 0\cr 0}\right).\eqno (6)$$
Now we can write up the Lagrangian of the simplest $SO(3)\rightarrow A_4$
theory:
$${\cal L}=-{1\over 4}F^{\mu\nu}_iF^i_{\mu\nu}+(D^{\mu}\Phi)^{\star}(D_{\mu}
\Phi)-\lambda (\Phi_0^2+...+\Phi_6^2-1)^2.\eqno (7)$$
Using (6) and (7) together we are able to study this ``exotic'' non-Abelian
discrete gauge theory. 
\section{Conclusions}
In our paper we have studied the $SO(3)$ gauge theory. We have developed a
general method to construct $SO(3)$-invariant polynomials and have given a
systematical description of the possible symmetry violations in the $SO(3)$
theory. Our results are important because it is possible that a general
discrete gauge theory in two spacetime dimensions possesses a strange field
theoretical symmetry, the so-called quantum symmetry [2],[3]. By the aid of
explicit examples like the $SO(3)\rightarrow A_4$ model we can study this
question very effectively. 
\section{References}
\hskip0.21in [1] M.G. Alford, J. March-Russel, F.Wilczek: {\it Discrete quantum hair on black holes and the non-Abelian Aharonov-Bohm effect},in: Nucl. Phys. B
{\bf B 337}, no. 3 (1990), 695-708.

[2] F. A. Bais, P. van Driel, M.de Wild Propitius: {\it Anyons in discrete
gauge theories with Chern-Simons terms}, in: Nucl. Phys., {\bf B893}, (1993),
547.

[3] F. A. Bais, P. van Driel, M.de Wild Propitius: {\it Quantum symmetries in 
discrete gauge theories}, in: Phys. Lett., {\bf B280}, (1992), 63-70.
   
\newpage
The following table shows the stabilizer subgroups of the group $SO(3)$ in an
arbitrary $2j+1$ dimensional irreducible representation.

The representation of weight $j$ contains the following stabilizer subgroups
systematically: 
\begin{itemize}
\item if $j$ is even and $j\geq 4$, then: 
$${\bf 1}, {\bf Z}_2\oplus{\bf Z}_2, {\bf Z}_3,...,{\bf Z}_j,
D_3,...D_j, O(2), SO(3),$$

\item if $j$ is odd and $j\geq 5$, then: 
$${\bf 1}, {\bf Z}_2\oplus{\bf Z}_2, {\bf Z}_3,...,
{\bf Z}_j, D_3,...,D_j, SO(2), SO(3),$$

\end{itemize}
Beyond these the not systematical groups  are showed on {\bf Table 1}.

The higher dimensional representations one can simple list:
\begin{itemize}
\item if $j\geq 30$ and is even the stabilizer subgroups in the $2j+1$ dimension
al representations are:
$${\bf 1}, {\bf Z}_2\oplus{\bf Z}_2, {\bf Z}_3,...,{\bf Z}_j,
D_3,...,D_j, A_4, A_5, S_4, O(2), SO(3),$$

\item if $j\geq 31$ and is odd then we get:
$${\bf 1}, {\bf Z}_2\oplus{\bf Z}_2, {\bf Z}_3,...,{\bf Z}_j,
D_3,...,D_j, A_4, A_5, S_4, SO(2), SO(3).$$
\end{itemize}
\newpage   
}
\centerline{
\vbox{\offinterlineskip
\halign{\strut
\vrule\hfil # \hfil &
\vrule\hfil # \hfil &
\vrule\hfil # \hfil &
\vrule\hfil # \hfil &
\vrule\hfil # \hfil &
\vrule\hfil # \hfil &
\vrule\hfil # \hfil &
\vrule\hfil # \hfil &
\vrule\hfil # \hfil &
\vrule\hfil # \hfil &
\vrule\hfil # \hfil\vrule\cr
\noalign{\hrule}
$\dim\rho$ & 1 & 3 & 5 & 7 & 9 & 11 & 13 & 15 & 17 & 19\cr
\noalign{\hrule}
    H    &$ SO(3)$ & $SO(2)$ & ${\bf Z}_2\oplus{\bf Z}_2$ & ${\bf 1}$ & $S_4$ &
 &
$A_4$ & $A_4$ & $S_4$ & $A_4$\cr
 &  & $SO(3)$ & $O(2)$ & $A_4$ &  & & $S_4$  & & & $S_4$\cr
 &  &  & $SO(3)$ & ${\bf Z}_3$ & & & $A_5$ & & & \cr
 &  &  &  & $D_3$ &  & & & & & \cr
 &  &  &  & $SO(2)$ &  & & & & & \cr
& & & & $SO(3)$ & & & & & & \cr
 \noalign{\hrule}
 }}}
 \vskip 0.3 in
 \centerline{
 \vbox{\offinterlineskip
 \halign{\strut
 \vrule\hfil # \hfil &
\vrule\hfil # \hfil &
\vrule\hfil # \hfil &
\vrule\hfil # \hfil &
\vrule\hfil # \hfil &
\vrule\hfil # \hfil &
\vrule\hfil # \hfil &
\vrule\hfil # \hfil &
\vrule\hfil # \hfil &
\vrule\hfil # \hfil &
\vrule\hfil # \hfil \vrule\cr
\noalign{\hrule}
$\dim\rho$ & 21 & 23 & 25 & 27 & 29 & 31 & 33 & 35 & 37 & 39 \cr
\noalign{\hrule}
    H    & $A_4$ & $A_4$ & $A_4$ & $A_4$ & $A_4$ & $A_4$ & $A_4$ & $A_4$ & $A_4$
 & $A_4$ \cr
         & $S_4$ &     & $S_4$ & $S_4$ & $S_4$ & $S_4$ & $S_4$ & $S_4$ & $S_4$ &
 $S_4$ \cr
         & $A_5$ &     &     &     &     & $A_5$ & $A_5$ &     & $A_5$ &     \cr
\noalign{\hrule}
}}}
\vskip 0.3 in
\centerline{
\vbox{\offinterlineskip
\halign{\strut
\vrule\hfil # \hfil &
\vrule\hfil # \hfil &
\vrule\hfil # \hfil &
\vrule\hfil # \hfil &
\vrule\hfil # \hfil &
\vrule\hfil # \hfil &
\vrule\hfil # \hfil &
\vrule\hfil # \hfil &
\vrule\hfil # \hfil &
\vrule\hfil # \hfil &
\vrule\hfil # \hfil \vrule\cr
\noalign{\hrule}
$\dim\rho$ & 41 & 43 & 45 & 47 & 49 & 51 & 53 & 55 & 57 & 59 \cr
\noalign{\hrule}
    H    & $A_4$ & $A_4$ & $A_4$ & $A_4$ & $A_4$ & $A_4$ & $A_4$ & $A_4$ & $A_4$
 & $A_4$ \cr
         & $S_4$ & $S_4$ & $S_4$ & $S_4$ & $S_4$ & $S_4$ & $S_4$ & $S_4$ & $S_4$
 & $S_4$ \cr
         & $A_5$ & $A_5$ & $A_5$ &     & $A_5$ & $A_5$ & $A_5$ & $A_5$ & $A_5$ &
     \cr
\noalign{\hrule}
}}}
\vskip 0.3 in
\centerline{{\bf Table 1.}} 
\centerline{The not-systematical stabilizer subgroups in the low dimensional}
\centerline{irreducible representations of the group $SO(3)$} 
\end{document}